\documentstyle[floats,multicol,aps,epsf,prl]{revtex}

\begin{document}
\draft
\twocolumn[\hsize\textwidth\columnwidth\hsize\csname @twocolumnfalse\endcsname
\def\btt#1{{\tt$\backslash$#1}}
\title{ Flow Induced Organization and Memory of a Vortex Lattice}
\author{Z. L. Xiao and E. Y. Andrei}
\address{Department of Physics and Astronomy, Rutgers University, 
Piscataway, New Jersey 08855}
\author{M. J. Higgins }
\address{NEC Research Institute, 4 Independence Way, Princeton, 
	 New Jersey 08540}
\maketitle
\begin{abstract}
{ We report on experiments probing the  evolution of a vortex state in
response to a  driving current in 2H-NbSe$_2$ crystals. 
By following the vortex motion with fast  transport measurements  we
find  that the current 
enables the system to reorganize and access new configurations. During
this process the system exhibits a long-term memory: 
if the current is turned off the vortices freeze in place  remembering their 
prior motion. When the current is restored the motion resumes where it
stopped.  The experiments provide evidence
for a dynamically driven structural change of the vortex lattice and a
corresponding dynamic phase diagram  that contains a previously
unknown regime where the critical current can be either $increased$ or 
$decreased$  by applying an appropriate driving current. }
 \end{abstract}
]

Vortices in a 
type II superconductor exhibit remarkably rich and sometimes
surprising collective behavior. 
The underlying physics is a delicate balance between two competing
forces: one is the mutual interaction between vortices which promotes
order and the other is pinning, caused by randomly distributed
material defects, which favors disorder \cite{1,2,3}. 
A striking example of this competition is the "peak effect" - a sudden
increase in the critical current just below the superconducting
transition temperature, $T_c$, attributed to a violent structural
reorganization when pinning overcomes the ordering influence of
interactions  \cite{3,4,5,6}.  Another outcome of this balance is that the
vortices are subject to an energy landscape with many metastable
states leading to history dependence and nonlinear behavior  \cite{4,5,6,7,8,9,10,11,12,13,14}. 
For example a vortex array prepared by field cooling, where the
magnetic field is applied before cooling through  $T_c$ ,  becomes trapped
in a metastable state that has a higher critical current and is more
disordered than one prepared by zero field cooling - where the field
is applied after cooling  \cite{4,5}. Setting the vortices in motion with an
external current releases them to explore the energy landscape and to
access new states  \cite{7,8,9}, but thus far the dynamics of this process was inaccessible to experiments. 
\par To probe the vortex  dynamics we have devised a technique that
 uses fast current ramps and pulses to track the
evolution of a  vortex lattice as it unpins and starts
moving. We find that when the vortices are set in motion 
 they organize into a new state whose properties are 
determined by the driving current.  The experiments lead to a  dynamic phase diagram 
containing four  regimes as a function of increasing current: 
A) pinned vortex array at the lowest currents; B) at higher currents a
region of  motional reorganization and repinning where the vortices
settle into a state whose critical current is equal to the driving
current; C) at still higher currents channels of ordered moving
vortices cut through domains of disordered stationary ones;
 D)  at the highest currents all the
vortices unpin and form an ordered state with low critical current. 

The vortex motion is detected by measuring the longitudinal  voltage 
across the sample {\it V= vBl}, where {\it v} is the vortex
velocity, $B$ the magnetic field  and $l$ the
distance between the voltage contacts. 
 As long as the vortex array is pinned the current flow is
dissipationless and only when a critical current ${\it I_{c}}$  is
exceeded, unpinning it, will a voltage appear. It was first shown by
Larkin and Ovchinnikov  \cite{3} that for weak pinning superconductors ${\it
I_{c}}$  is a measure of the degree of order in the vortex system. In
this case, as was confirmed by recent SANS  \cite{15} and transport
measurements  \cite{5}, the critical current density is inversely
proportional to the size of an ordered vortex domain and ${\it I_{c}}$
can be used, as was done here, to quantify the degree of vortex order.

The data  were acquired on a Fe (200ppm) doped single
crystal of 2H-NbSe$_2$ of  size of (3$\times $1$\times $0.015) mm$^3$ and
$T_c$ = 5.95K. Our measurements employed a  standard 
 four probe technique with low resistance
contacts made with Ag$_{0.1}$In$_{0.9}$ solder. The current
ramps and pulses  were generated with a  Wavetek 164 and the 
 voltage response was measured with  a fast (INA103) amplifier. 
 The magnetic field was kept
along the c axis of the sample and the current in the a-b
plane. All the data were taken after the sample is zero field cooled (ZFC).
 Previous experiments on similar samples revealed novel
types of nonlinear behavior including frequency memory and long glassy
relaxation times that were observed in the lower part of the peak
effect region  \cite{16}. In the present experiments we find that this is
also the region where the value of the critical current changes during the measurement.

\begin{figure}[btp]
\epsfxsize=3.5in
\epsfysize = 3.0in
\epsfbox{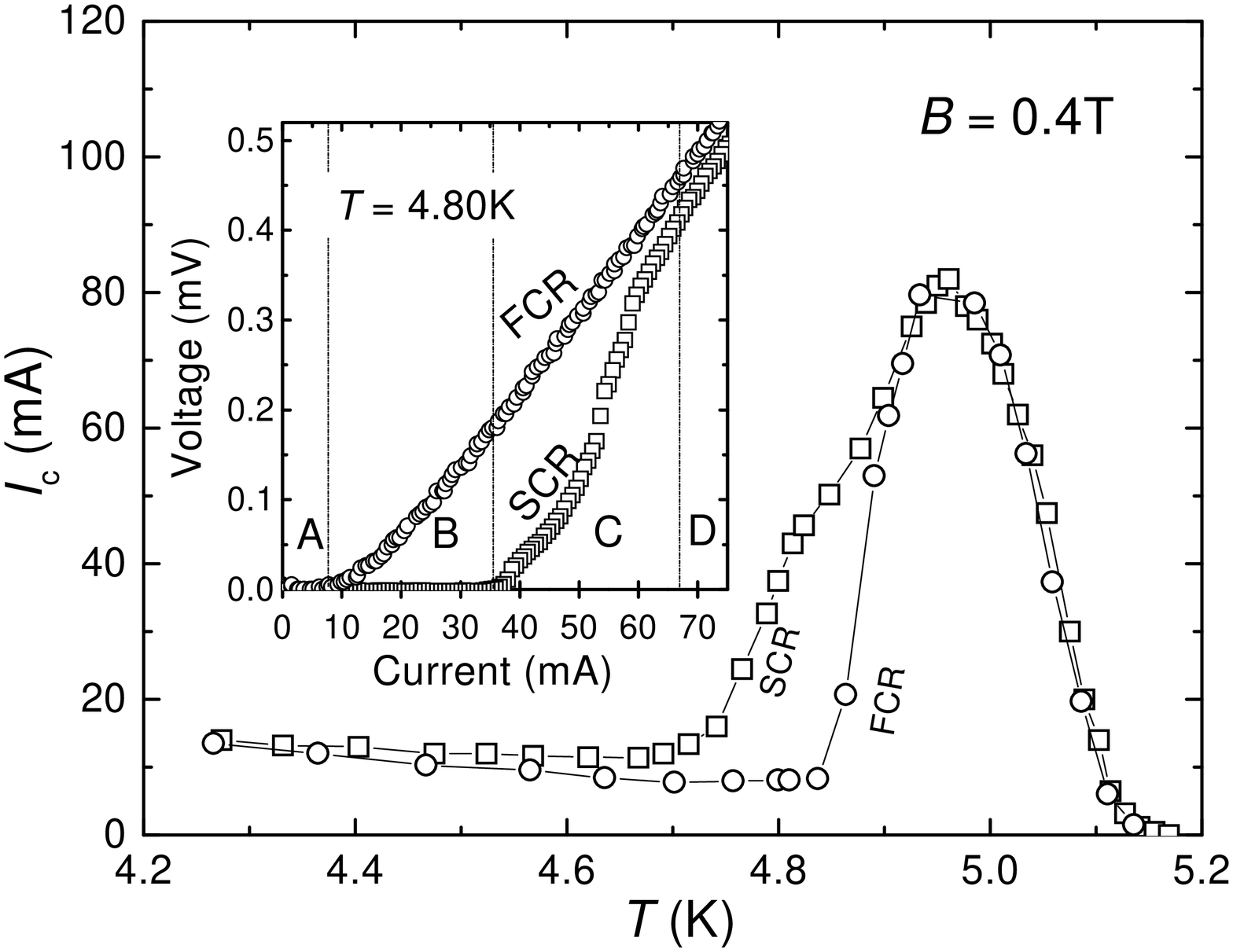} 
\caption{Temperature dependence of ${I_c}$  detected by 
slow (SCR) and fast (FCR) current ramps. Inset: $ I-V$ curves in response to  SCR (1mA/s) and FCR (200A/s). The labels  A,B,C,D illustrate the four regions of the dynamic phase diagram. }
\label{fig:Fig1}
\end{figure}

Fig.1 shows the response  to two types of current ramps, 
fast (FCR -ramping rate of 200A/s), and slow current ramps (SCR
-ramping rate of 1mA/s). The critical current 
 was defined as the current at which the voltage reaches 1$\mu $V. 
The inset in Fig.1, which shows the current-voltage ($I-V$)
characteristics for both ramps, illustrates that the response  depends on the speed of measurement. 
In particular the critical current $ \it {I_{cf}}$,
 obtained with the FCR, is lower than $\it{ I_{cs}}$ , the value
obtained for the SCR.   The temperature dependence of $\it{ I_{cs}}$
and $\it{ I_{cf}}$ is shown in the main panel of Fig.1. 
 A strong peak effect as reported in Ref.[5] is seen in both data
sets. The fact that $\it{ I_{cf} \leq I_{cs}}$,
 for all data points means that the  ZFC  state probed by 
 the FCR is always more ordered than the SCR state, leading to the
conclusion  that
during the SCR the vortex array has time to reorganize into a more
strongly pinned state.  This organization   does not however produce a voltage
signal in the SCR measurements because it decays too fast to be
resolved. 

To observe the reorganization directly we carried out fast measurements with
 2$\mu $sec temporal resolution, which enabled us to follow the decay of the 
voltage response to a current step $ \it{I> I_{cf}}$  as shown in Fig.2. 
\begin{figure}[btp]
\epsfxsize=3.5in
\epsfysize = 2.5in
\epsfbox{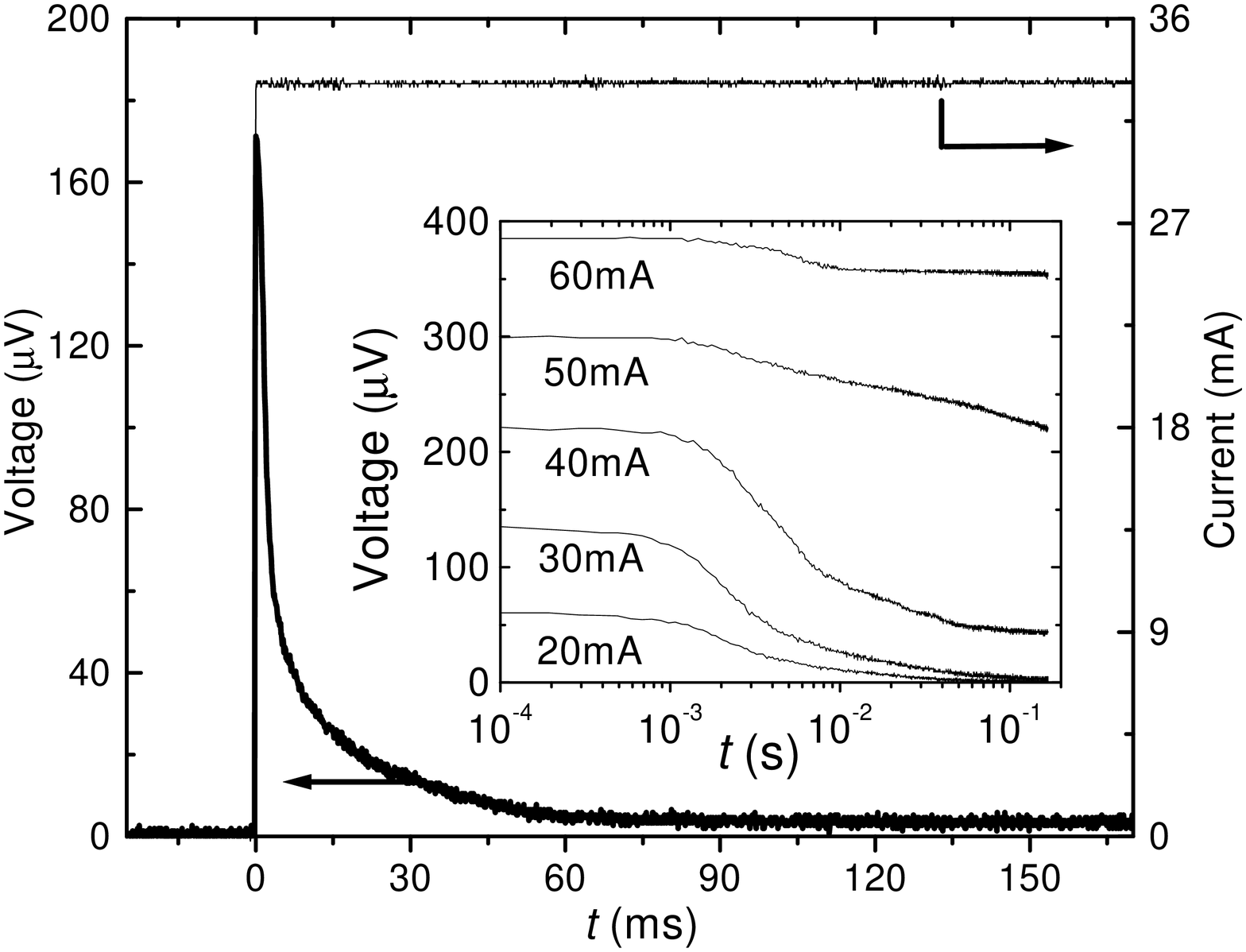} 
\caption{Time dependence of response to a current step. 
The inset shows the  response to steps of different amplitudes. }
\label{fig:Fig2}
\end{figure}
 We note that  upon applying  the current step  the vortex
response is  instantaneous  as expected of  essentially massless
objects  \cite{17} when they respond elastically. 
As time progresses the voltage decreases despite the fact that the
driving current is kept constant.  After a sufficiently long waiting time the voltage decays to zero for
driving currents in the range ${\it I_{cf}<I< I_{cs}}$.  The decay
is also observed for driving currents  in
excess of ${\it I_{cs}}$  indicating that the moving vortex array
organizes itself into less ordered state even though it is not coming
to a full stop.
In either case the  organization proceeds in two steps. 
On long time scales there is a clear decay which exhibits a stretched exponential time dependence
exp(-t/$\tau )^\alpha$ where both the characteristic time scale, $\tau
$ and the exponent $\alpha $ depend on current and temperature. 
This type of relaxation is typical of the dynamics observed in
glasses. By contrast, on short
time scales  the response is almost constant 
showing no measurable decay. 
This is consistent with the  observation of Ref. [8] that the 
vortex pattern does not change instantaneously. 
It is this delay before the decay sets in that makes it possible to
observe the pristine vortex state - before it has time to reorder - 
by using a fast current ramp technique. 

During the reorganization a striking long-term memory is observed. 
A typical result is shown in 
 Fig.3(a) where we plot the response to a
current pulse of amplitude $I$. 
 When the current is interrupted during the decay, the voltage drops
to zero instantaneously. Subsequently, when the current is restored
after a waiting time with zero current, 
the voltage decay resumes where it stopped the moment  the  current was
turned off. 
The waiting times for which the lattice remembers its state of motion
prior to being stopped can be very long (we checked waiting times of
up to 20 hours) provided the field and temperature remain unchanged. 
Other effects associated with quenching of the vortex motion where
observed in experiments on the structure  \cite{7,18,19} and transport
properties  \cite{20}. 

 	Long-term memory occurs in systems that can be trapped
indefinitely in a metastable state. In  the case of a vortex array
this implies  that the energy scale
of the pinning potential is much larger than that of thermal
fluctuations. The current lowers the barriers of the pinning potential 
and sets the vortices in motion allowing them to explore new states
and eventually settle into a more strongly pinned
 state.  It is thus the applied current rather than the
temperature that drives the
 system towards equilibrium.  How robust are 
these metastable states or equivalently how large can the current be before the vortices become 
untrapped? To  explore the interplay between  current and  pinning  we carried out the 
experiment shown in Fig.3(b). The ZFC vortex lattice 
is driven with a current 
pulse of amplitude $I$ for 1.6msec during which the
response decays  correspondingly. 
The current is then suddenly reduced to a lower value $I_0$ for 6msec during which the 
voltage decays at a different rate, and subsequently it is restored to its original value. 
Plotting $\delta $V, the voltage decay during the reduced current
pulse, as a function of  $I_0$, 
 we find that there 
is indeed a finite value of  $I_0$ 
 below which the system does not evolve with time and retains its memory. 
This value of the current coincides with the critical current of the vortex lattice 
at the moment at which the reduced current is applied.

\begin{figure}[btp]
\epsfxsize=3.5in
\epsfbox{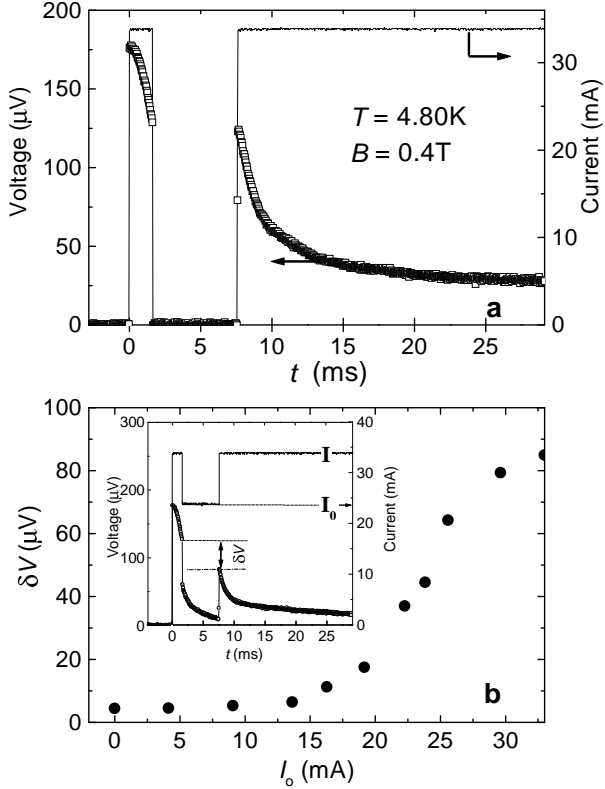} 
\caption{Memory of a vortex array. (a) During the decay  the current
is removed and later it 
is restored to its initial value.
 The lines and the open circles are the current and the response 
 respectively. (b) $\delta V$ as a function of $I_0$. The definitions of
$\delta V$ and  $I_0$ are shown in inset.}
\label{fig:Fig3}
\end{figure}

The memory can be a powerful tool in the study of vortex dynamics. 
Most probes of the vortex structure require long times to
produce one data point or a picture, typically 45 seconds in scanning
Hall probe microscopy  \cite{21} and 15 minutes in SANS  \cite{7}. 
It would be impossible to observe the motional reorganization with
these techniques. But as our experiments have shown,  
 the vortex structure can be frozen at any moment during its
motion by removing the driving current. 
The quenched vortex system maintains its structure and can then be
probed directly. We used this method here to observe the vortex reorganization
by  quenching the  system at various times during its motion and
measuring its critical current  with the FCR method.

\begin{figure}[btp]
\epsfxsize=3.5in
\epsfysize = 4.0in
\epsfbox{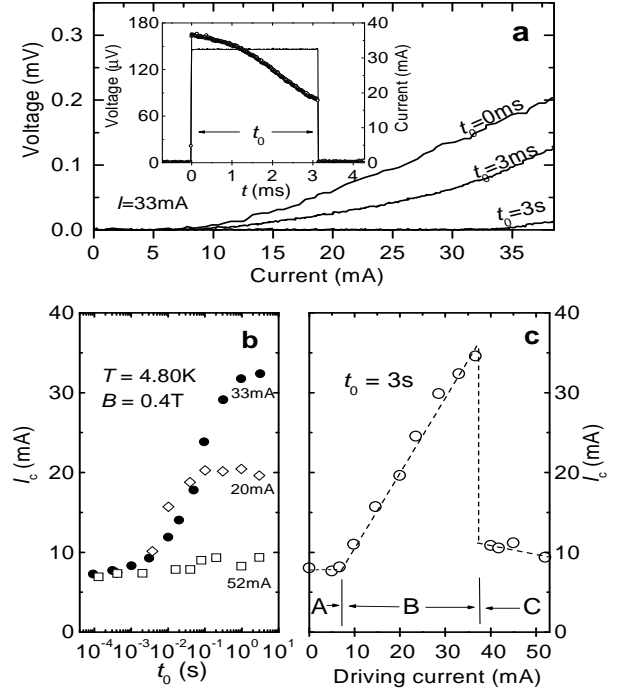} 
\caption{(a) $ I-V$ curves obtained with a fast current ramp following decay periods 
$t_0$=0, 3ms and 3s with current ${I}$=33mA. The definition of $t_0$
 is shown in the inset. (b) Evolution of critical current with time in the presence of driving currents 
 ${I}$ = 20, 33, and 52mA. (c) Data for  ${I_c}$  obtained at $t_0$ =
 3s and various driving currents.
The symbols A, B, C represent the three regimes of the dynamic phase diagram. The dashed line is a guide to the eye.}

\label{fig:Fig4}
\end{figure}

 As an example, we show in Fig.4(a) the response to an FCR at decay
 times $\it {t_0}$ = 0, 3ms and 3s. 
Clearly, $\it{I_c}$ increases with increasing decay time. 
In Fig.4(b) we show the dependence of $\it{I_c}$ on decay time for
 driving currents $\it{I}$ = 20, 33 and 52mA. For current amplitudes
 $\it{I_{cf}<I\leq I_{cs}}$, the critical current increases with
 decay time and saturates at a value which is approximately equal to
 the driving current.  In  Fig.4(c) we show the dependence of $\it
 { I_{c}}$  on $\it{I}$  following a long decay time of $\it {t_0}$
 = 3s. As long as $\it{I_{cf}<I\leq I_{cs}}$ 
 we find that $\it{ I_{c}\sim I}$, in other words the system  can be prepared
 with any desired critical current in this range  simply by 
priming it with the appropriate driving current. For $\it{I>I_{cs}}$, 
 there is an abrupt drop in $\it{ I_{c}}$  to a value only slightly above that in 
the ordered state. We interpret this drop in terms of the formation of a
 channel of unpinned and motionallay ordered vortices. The fact that critical 
current is slightly higher than that of the ZFC state may indicate that, as predicted by theory  \cite{10,11}, the motionally ordered state resembles a smectic rather than a crystal.  

We now show that for  ${\it I>I_{cs}}$ our data is consistent with channel formation 
 and  estimate the number of vortices in  channels.
First we interpret the information contained in the $ I-V$ curves in terms of 
vortex motion: ${\it I_{c}}$  gives the force required to set the
 vortices in motion and the cordal resistance, defined as 
$\it {R_c =V/(I -I_{c})}$, 
measures the number of moving vortices in response to a current $I$. 
When all the vortices are moving $R_c$ takes the maximum value,
 given by the Bardeen-Stephen 
resistance  \cite{22} $R_{BS}=R_n(\Phi _0/H_{c2}$)n, where $R_n$ is the normal 
state resistance, $\Phi _0$ the fundamental unit of flux, 
$ H_{c2}$ the upper critical field and $ n$ is the 
vortex density. If only a fraction of the vortices move, $R_c$ is reduced
 proportionately. We note that for the FCR data in the inset of Fig.1 
the cordal resistance is close to the maximum value, $R^f_c$= 8.75m$\Omega $ = 0.94$R_{BS}$  
which means that essentially all the vortices move.
By contrast the $ I-V$ curve obtained with the SCR for the same initial vortex
 state is different: it has a much higher critical current and once
 the vortex motion is detectable the slope increases with current. 
In this case the number of moving vortices depends on the current and can be estimated by using the result of Fig.4(c) that  for 
$\it {I> I_{cs}}$ , the moving vortex array is in a state with
 critical current $\sim I_{cf}$.  Therefore, 
the ratio of voltage response obtained with the SCR to that obtained
 with the FCR $\it{ V_s(I)/V_f(I)= R_c^s /
 R_c^f}$  is equal to the ratio of moving vortices in the two
 measurements. 
Since in the FCR case the entire  array is moving 
 $\it{ V_s(I)/V_f(I)}$ is in fact a direct  measure the fraction of
 moving vortices. This result leads to the conclusion that for 
 $\it{ I_{cs}<I<2 I_{cs}}$  not all the vortices are participating
 in the motion and those that do move form channels in which the
 vortices are ordered (because the critical current is close to $\it
 {I_{cf}}$ ), consistent with our assumption. 
The fraction of moving vortices in the SCR increases gradually from
 zero at $I= \it{ I_{cs}}$  to 1 when  the
 channels have engulfed the entire sample forming an 
ordered state. 

Based on these results we  identify four regions in the dynamic phase
diagram as illustrated in Fig.1. In region A where $\it{I< I_{cf}}$  , 
the vortex lattice stays pinned and ordered. Region B, where 
$\it{ I_{cf} <I< I_{cs}}$, is the regime of motional reorganization and 
repinning. Here the response decays with time as parts of the lattice break 
loose settling into stronger pinned configurations until the motion comes to a halt when the vortices reach a state with $\it{ I_{c}}$  equal to the driving 
current.
 This state contains at least one contiguous region of disordered vortices that provides a low resistance current path along the sample. Region C, 
where $\it {I_{cs} <I<2 I_{cs}}$  , is the plastic flow regime in which 
channels of moving ordered vortices cutting across the sample are embedded in domains of disordered stationary vortices. In region D, $\it{  I>2I_{cs}}$ , 
the entire vortex array unpins and becomes motionally ordered. 

We conclude that the current assists the system in choosing a configuration in the energy landscape formed by the competition between interactions and pinning. 
The  current assisted rearrangement can also account for the fact that  the ZFC 
vortex state has the lowest critical current: as the field is ramped
up in the superconducting  state, it is initially expelled from the
interior of the sample by the screening currents. 
The vortices can penetrate only when the screening current density
exceeds the critical value, and therefore the configuration with the
lowest critical current- the ordered state  \cite{23} -will be the
first to penetrate the  sample. 
For the experiments discussed here the typical ordered domain
in the ZFC state contains $\sim 10^3$ vortices. 
Once the vortices are inside the sample an externally applied current assists
them in finding a lower energy configuration by adapting to the
pinning potential and becoming more disordered. 
With increasing current, as the pinning barriers are effectively
lowered, larger portions of the pinscape can be explored resulting in
progressively more strongly pinned states. 
Until at $\it{  I>I_{cs}}$  a channel 
forms across the sample in which no barriers are encountered 
allowing free passage of moving, ordered vortices. 
The motional reorganization does not occur immediately
after the array is set in motion but after a delay during
which it maintains its prior configuration. 
This initial hesitation together with the long term memory makes it
possible to directly  probe these processes  with
transport measurements and in principle with other techniques as
well. 
 
We thank N. Andrei, R. Chitra, W. Henderson, Y. Paltiel, A. Rosch, R. Ruckenstein, S. Bhattacharya and E. Zeldov for useful discussion. The work at Rutgers was supported by NSF-DMR and by DOE.

\end{document}